\documentclass[journal, two column]{IEEEtran}

\usepackage{cite}
\usepackage{graphicx}
\usepackage{psfrag}
\usepackage{subfigure}
\usepackage{url}
\usepackage{stfloats}
\usepackage{amsmath}
\usepackage{array}
\usepackage{amssymb}
\usepackage{setspace}
\usepackage{algorithmicx}
\usepackage[ruled]{algorithm}
\usepackage{algpseudocode}
\usepackage{algpascal}
\usepackage{algc}

\begin{document}

\title{Low-Complexity QL-QR Decomposition Based Beamforming Design for Two-Way MIMO Relay Networks}
\author{Wei Duan,~\IEEEmembership{Student Member,~IEEE} Wei Song and Moon Ho Lee,~\IEEEmembership{Senior Member, IEEE}
\thanks{Wei Duan,  and Moon Ho Lee are with the Institute of Information and Communication, Chonbuk National University, 664-14 Deokjin-dong, Jeonju 561-756, Republic of Korea.(e-mail:
sinder@live.cn, moonho@jbnu.ac.kr)}
\thanks{Wei Song is with College of
Information Technology, Eastern liaoning university,  Dandong,
118003, P.R.China. (e-mail: sw658@hotmail.com)}
} \maketitle

\begin{abstract}
In this paper, we investigate the optimization problem of joint
source and relay beamforming matrices for a two-way
amplify-and-forward (AF) multi-input multi-output (MIMO) relay
system. The system consisting of two source nodes and two relay
nodes is considered and the linear minimum mean-square-error
(MMSE) is employed at both receivers. We assume individual relay
power constraints and study an important design problem, a
so-called determinant maximization (DM) problem. Since this DM
problem is nonconvex, we consider an efficient iterative algorithm
by using an MSE balancing result to obtain at least a locally
optimal solution. The proposed algorithm is developed based on QL,
QR and Choleskey decompositions which differ in the complexity and
performance. Analytical and simulation results show that the
proposed algorithm can significantly reduce computational
complexity compared with their existing two-way relay systems and
have equivalent bit-error-rate (BER) performance as the singular
value decomposition (SVD) based on a regular block diagonal (RBD)
scheme.

\end{abstract}

\section{Introduction}

Recently, wireless relay networks have been the focus of a lot of
research because the relaying transmission is a promising
technique which can be applied to extend the coverage or increase
the system capacity. There are various cooperative relaying
schemes have been proposed, such as amplify-and-forward (AF)
\cite{b1} and \cite{c2}, decode-and-forward (DF) \cite{b2},
denoise-and-forward (DNF) \cite{b3}, and compress-and-forward (CF)
\cite{b4} cooperative relaying protocols. Among these approaches,
AF is most widely used due to without detecting the transmitted
signal. Therefore, an AF relay scheme requires a less processing
power at the relays compared to other schemes.

In one-way relaying (OWR) approach, to completely exchange
information between two base stations, four time slots are
required in uplink (UL) and downlink (DL) communications, which
leads to a loss of one-half spectral resources \cite{b5}. In order
to solve this problem, a two-way relaying approach has been
considered in \cite{b6}, \cite{b7}, and \cite{b8}. In a typical
two-way relaying scheme, the communication is completed in two
steps. First, the transmitters send their symbols to two relays,
simultaneously. After receiving the signals, each relay processes
them based on an efficient relaying scheme to produce new signals.
Then the processed signals are broadcasted to both receiver nodes.

Multi-input multi-output (MIMO) relay systems have been
investigated in [10]--[13]. It is shown that, by employing
multiple antennas at the transmitter and/or the receiver, one can
significantly improve the transmission reliability by leveraging
spatial diversity. Relay precoder design methods have been
investigated in \cite{b13}\raisebox{0.1mm}, \cite{b14},
\cite{b15}. A problem in designing optimal beamforming vectors for
multicasting is challenging due to its nonconvex nature. In
\cite{b13}, the authors propose a transceiver precoding scheme at
the relay node by using zero-forcing (ZF) and MMSE criteria with
certain antenna configurations. The information theoretic capacity
of the multi-antenna multicasting is studied in \cite{b14}, along
with the achievable rates using lower complexity transmission
schemes, as the number of antennas or users goes to infinity. In
\cite{b15}, the authors propose an alternative method to
characterize the capacity region of two-way relay channel (TWRC)
by applying the idea of rate profile.

Joint optimization of the relay and source nodes for the MIMO TWRC
have been studied in \cite{b8}, \cite{b16}. In \cite{b8}, the
authors develop a unified framework for optimizing two-way linear
non-regenerative MIMO relay systems and show that the optimal
relay and source matrices have a general beamforming structure.
The joint source node and relay precoding design for minimizing
the mean squared error in a MIMO two-way relay (TWR) system is
studied in \cite{b16}.

Since singular value decomposition (SVD) and/or generalized SVD
(GSVD) are widely used to find the orthogonal complement to solve
an optimization problem \cite{c2}, \cite{b8}, \cite{b15},
\cite{b19}, but their computational complexities are extremely
high. In order to reduce the complexity, the SVD can be replaced
with a less complex QR decomposition \cite{f1} in this work.
However, this approach leads to degrading the BER performance. In
addition, it is difficult to realize in TWRC. In this paper, we
investigate the joint source and relay precoding matrix
optimization for a two way-relay amplify-and-forward relaying
system where two source nodes and two relay nodes are equipped
with multiple antennas. Also, in order to apply the QL/QR
decomposition to the TWRC, we design a three part relay filter.
Compared with existing works such as [9]--[14], the contributions
of this paper can be summarized as follows. Firstly, we
investigate a two-way MIMO relay system using the criteria which
minimizes an MSE of the signal waveform estimation for both two
source nodes. We prove an optimal sum-MSE solution can be obtained
as the Winer filter while signal-to-noise-ratio (SNR) at both
source nodes are equivalent \cite{b17} which leading to an MSE
balancing result. Secondly, we propose a new cooperative scenario,
i.e., the QL-QR compare with the Choleskey decomposition which
significantly reduces the computational complexity of the optimal
design. In this proposed design, the channels of its left side are
decomposed by the QL decomposition while those of its right side
factorized by the QR decomposition. And the equivalent noise
covariance is decomposed by the Choleskey decomposition. We also
design the three part relay filter, which is comprised by a left
filer, a middle filter, and a right filter, to efficiently combine
two source nodes and the relay nodes. By these approaches, the
received signals at both two source nodes are able to be redeemed
as either lower or upper triangular matrices. Stemming from one of
the properties of triangular matrices such that their determinant
is identical to the multiplication of their eigenvalues, we are
able to straightforwardly solve the optimization problem as a
determinant maximization problem. Also, we can obtain the BER
performance equivalent to that of SVD-RBD scheme.

The rest of this paper is organized as follows. Section II
describes a system model of the TWRC and raises a sum-MSE problem.
In Section III, we propose an iterative QL-QR algorithm and a
joint optimal beamforming design. In Sections IV, we discuss the
computational complexity of an efficient channel model. The
simulation results are presented to show the excellent performance
of our proposed algorithm for the TWRC in Section V. Section VI
concludes this paper.

\emph{Notations:} $\mathbf{A}^*$ and $\mathbf{A}^T$ denote the
conjugate and the transpose of a matrix $\mathbf{A}$,
respectively. $\mathbf{d}(.)$ denotes a diagonal matrix and an
$N\times N$ identity matrix is denoted by $\mathbf{I}_N$.
$\mathrm{E}(.)$ stands for the statistical expectation and
$(.)^{H}$ denotes a Hermitian transpose of a given matrix. $tr(.)$
and $R(.)$ denotes matrix trace and the range of a matrix.

\begin{figure}
\begin{center}
\includegraphics [width=60mm,height=34mm]{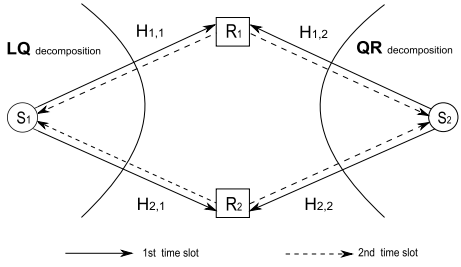}
\caption{\label{1} Proposed LQ-QR Amplify-and-Forward MIMO TWR
system.}
\end{center}
\end{figure}

\section{System Model and Sum-MSE}
 We consider a TWRC consisting of two source nodes
$\mathbf{S}_{1}$ and $\mathbf{S}_{2}$, and two relay nodes
$\mathbf{R}_{1}$ and $\mathbf{R}_{2}$  as shown in Fig. \ref{1}.
The source and relay nodes are equipped with \emph{M} and \emph{N}
antennas, respectively. We adopt the relay protocol with two time
slots introduced in \cite{b13}. In the first time slot, the
information vector $x_{i}\in{\mathbb{C}^{M\times{1}}}$ is linearly
processed by a precoding matrix
$\mathbf{V}_{i}\in{\mathbb{C}^{M\times{M}}}$ and then be
transmitted to the relay nodes. The received signals at
$\mathbf{R}_{i}$, $i\in{\{1,2\}}$, can be expressed as
\begin{eqnarray}\label{t1}
&&y_{R_{1}}=\textbf{H}_{1,1}s_{1}+\textbf{H}_{1,2}s_{2}+n_{R_{1}}\notag\\
&&y_{R_{2}}=\textbf{H}_{2,1}s_{1}+\textbf{H}_{2,2}s_{2}+n_{R_{2}},
\end{eqnarray}
where $y_{R_{i}}\in{\mathbb{C}^{N\times1}}, i\in\{1,2\}$,
indicates the received signal vector,
$\textbf{H}_{i,j}\in{\mathbb{C}^{N\times{M}}}, i,j\in\{1,2\}$,
represents the channel matrix from source $j$ to relay $i$, as
shown in Fig.1, $s_{i}\in{\mathbb{C}^{M\times1}}$ is the
transmitted symbol vector from $\mathbf{S}_{i}$, and
$n_{R_{i}}\sim{CN(0,\sigma^{2}_{R_{i}}\mathbf{I}_{N})}$ represents
the additive white Gaussian noise (AWGN) vector with zero mean and
variance $\sigma^{2}_{R_{i}}$ at relay node $i$. The term $s_{i}$
is subject to a power constraint,
$tr\{E(s_{i}s^{H}_{i})\}\leq{P_{i}}$ with
$tr\{E(x_{i}x^{H}_{i})\}\leq{\frac{P_{i}}{M}\mathbf{I}_{M}}$,
where $P_{i}$ is the transmit power at $\mathbf{S}_{i}$.

To find an appropriate power normalization vector $\rho_{R}$, we
express the total transmission power at the source node with
$\mathbf{V}_{i}$ as
\begin{eqnarray}\label{Power nomarlizeition}
tr\{\mathbf{V}_{1}\mathbf{V}^{H}_{1}\!+\!\mathbf{V}_{2}\mathbf{V}^{H}_{2}\}\!\!\!\!&=&\!\!\!\!tr\left\{\rho^{2}_{R}\left(\textbf{V}^{b}_{1}\left(\textbf{V}^{b}_{1}\right)^{H}+\textbf{V}^{b}_{2}\left(\textbf{V}^{b}_{2}\right)^{H}\right)\right\}\notag\\
&=&\!\!\!\!tr\left\{\rho^{2}_{R}\left(P_{1}+P_{2}\right)\right\}.
\end{eqnarray}
In this paper, we assume that each transmit antenna satisfies the
unity transmission power constraint. To satisfy the power
constraint, we propose the following power normalization vector
\begin{eqnarray}\label{Power nomarlizeition1}
\rho_{R}=1/\sqrt{P_{1}+P_{2}}.
\end{eqnarray}
In the second time slot, after power normalization, the relay node
$\mathbf{R}_{i}$ linearly amplifies $y_{R_{i}}$ with an
$N\times{N}$ matrix $\mathbf{F}_{i}$ and then broadcasts the
amplified signal vector $x_{R_{i}}$ to source nodes 1 and 2. The
signals transmitted from relay node $i$ can be expressed as
\begin{eqnarray}\label{R power}
x_{R_{i}}=\rho_{R}\textbf{F}_{i}y_{R_{i}}.
\end{eqnarray}
Using \eqref{t1} and \eqref{R power}, the received signal vectors
at $\mathbf{S}_{1}$ and $\mathbf{S}_{2}$ can be, respectively,
written as
\begin{eqnarray}\label{s2}
y_{1}&=&\textbf{H}^{T}_{1,1}\textbf{F}_{1}\textbf{H}_{1,2}s_{1}+\textbf{H}^{T}_{1,1}\textbf{F}_{1}\textbf{H}_{1,2}s_{2}
+\textbf{H}^{T}_{2,1}\textbf{F}_{2}\textbf{H}_{2,2}s_{1}\notag\\
&&+\textbf{H}^{T}_{2,1}\textbf{F}_{2}\textbf{H}_{2,2}s_{2}
+\textbf{H}^{T}_{1,1}\textbf{F}_{1}n_{R_{1}}
+\textbf{H}^{T}_{2,1}\textbf{F}_{2}n_{R_{2}}+n_{1}\notag\\
y_{2}&=&\textbf{H}^{T}_{1,2}\textbf{F}_{1}\textbf{H}_{1,1}s_{1}+\textbf{H}^{T}_{1,2}\textbf{F}_{1}\textbf{H}_{1,1}s_{2}
+\textbf{H}^{T}_{2,2}\textbf{F}_{2}\textbf{H}_{2,1}s_{1}\notag\\
&&+\textbf{H}^{T}_{2,2}\textbf{F}_{2}\textbf{H}_{2,1}s_{2}+\textbf{H}^{T}_{1,2}\textbf{F}_{1}n_{R_{1}}+\textbf{H}^{T}_{2,2}\textbf{F}_{2}n_{R_{2}}+n_{2},\notag\\
\end{eqnarray}
where ${\textbf{H}}^{T}_{i,j}$, $i,j\in\{1,2\}$, indicates the
$M\times{N}$ channel matrix from the relay node $i$ to the source
node $j$, and $n_{i}$, $i\in\{1,2\},$ is an $M\times 1$ noise
vector at $\mathbf{S}_{i}$.
We assume that the relay nodes perfectly know the channel state
information (CSI) of $\textbf{H}_{i,j}$. The relay node
$\mathbf{R}_{i}$ performs the optimizations of $\textbf{F}_{i}$
and $\mathbf{V}_{i}$, and then transmits the information to the
source nodes 1 and 2. Since source node $i$ knows its own
transmitted signal vector $s_{i}$ and full CSI, the
self-interference components in \eqref{s2} can be efficiently
cancelled. The effective received signal vectors are given by
\begin{eqnarray}\label{s3}
\widetilde{y}_{1}&=&\textbf{H}^{T}_{1,1}\textbf{F}_{1}\textbf{H}_{1,2}s_{2}+\textbf{H}^{T}_{2,1}\textbf{F}_{2}\textbf{H}_{2,2}s_{2}+\textbf{H}^{T}_{1,1}\textbf{F}_{1}n_{R_{1}}\notag\\
&&+\textbf{H}^{T}_{2,1}\textbf{F}_{2}n_{R_{2}}+n_{1}\notag\\
&=&\mathbf{\widetilde{H}}_{1}s_{2}+\widetilde{n}_{1},
\end{eqnarray}
\begin{eqnarray}\label{a}
\widetilde{y}_{2}&=&\textbf{H}^{T}_{1,2}\textbf{F}_{1}\textbf{H}_{1,1}s_{1}+\textbf{H}^{T}_{2,2}\textbf{F}_{2}\textbf{H}_{2,1}s_{1}+\textbf{H}^{T}_{1,2}\textbf{F}_{1}n_{R_{1}}\notag\\
&&+\textbf{H}^{T}_{2,2}\textbf{F}_{2}n_{R_{2}} +n_{2}\notag\\
&=&\mathbf{\widetilde{H}}_{2}s_{1}+\widetilde{n}_{2},
\end{eqnarray}
where
$\widetilde{\textbf{H}}_{1}=\textbf{H}^{T}_{1,1}\textbf{F}_{1}\textbf{H}_{1,2}+\textbf{H}^{T}_{2,1}\textbf{F}_{2}\textbf{H}_{2,2}$
and
$\widetilde{\textbf{H}}_{2}=\textbf{H}^{T}_{1,2}\textbf{F}_{1}\textbf{H}_{1,1}+\textbf{H}^{T}_{2,2}\textbf{F}_{2}\textbf{H}_{2,1}$
are the equivalent MIMO channels seen at source nodes
$\mathbf{S}_{1}$ and $\mathbf{S}_{2}$, respectively. The vectors
$\widetilde{n}_{1}=\textbf{H}^{T}_{1,1}\textbf{F}_{1}n_{R_{1}}+\textbf{H}^{T}_{2,1}\textbf{F}_{2}n_{R_{2}}+n_{1}$
and
$\widetilde{n}_{2}=\textbf{H}^{T}_{1,2}\textbf{F}_{1}n_{R_{1}}+\textbf{H}^{T}_{2,2}\textbf{F}_{2}n_{R_{2}}
+n_{2}$ are the equivalent noises at source node $\mathbf{S}_{1}$
and $\mathbf{S}_{2}$, respectively.

Due to the lower computational complexity, linear receivers are
applied at source node $i$ to retrieve the transmitted signals
sent from the other nodes. The estimated signal waveform vector is
given as
$\widehat{s}_{\overline{i}}=\mathbf{W}^{H}_{i}\widetilde{y}_{i}$,
where $\mathbf{W}_{i}$ is an $M\times{M}$ weight matrix, with
$\overline{i}=2$ for $i=1$ and $\overline{i}=1$ for $i=2$. From
\eqref{s3}, the MSE matrix of the signal waveform estimation
denoted by
$\mathbf{MSE}_{\overline{i}}=\mathrm{E}[(\widehat{s}_{i}-{s}_{i})(\widehat{s}_{i}-{s}_{i})^{H}]$,
which can be further written as
\begin{eqnarray}\label{mse}
\mathbf{MSE}_{i}&=&(\mathbf{W}^{H}_{i}\widetilde{\mathbf{H}}_{i}-\mathrm{I}_{M})(\mathbf{W}^{H}_{i}\widetilde{\mathbf{H}}_{i}-\mathrm{I}_{M})^{H}\notag\\
&&+\mathbf{W}^{H}_{i}\mathbf{C}_{n_{i}}\mathbf{W}_{i}
\end{eqnarray}
where
$\mathbf{C}_{n_{i}}=\textbf{H}^{T}_{i,i}\textbf{F}_{i}\textbf{F}^{H}_{i}\textbf{H}^{*}_{i,i}+\textbf{H}^{T}_{\overline{i},i}\textbf{F}_{\overline{i}}\textbf{F}^{H}_{\overline{i}}\textbf{H}^{*}_{\overline{i},i}+\mathbf{I}_{M}$
is the equivalent noise covariance. The sum-MSE of the two source
nodes in the proposed system model can be written as:
\begin{eqnarray}\label{s4}
\mathbf{MSE}_{sum}=\mathbf{MSE}_{1}+\mathbf{MSE}_{2}.
\end{eqnarray}
Note that the sum-MSE minimization criterion measures the overall
transmission performance of both the DL and the UL. Since the two
data streams are transmitted at different directions during the
two time slots are considered in the TWR network.

\section{Joint Source and relay Beamforming Design}

In this section, we develop an iterative QL-QR algorithm by using
the MSE balancing result. The QL-QR algorithm involves two steps,
i.e., the linear receiver matrix optimization and the joint source
and relay beamformer design.

\begin{figure}
\begin{center}
\includegraphics [width=115mm,height=55mm]{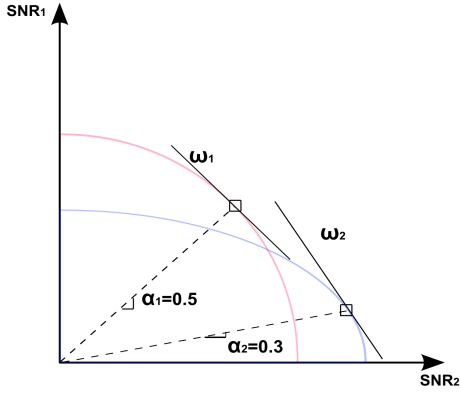}
\caption{\label{2} Examples of the SNR regions achieved in a TWRC
with two relays.}
\end{center}
\end{figure}

\subsection{Proposed Optimal Detector and Optimization Problem}
We would like to find the jointly optimal beamforming vectors
$\mathbf{W}_{i}$, $\mathbf{V}_{i}$ and $\mathbf{F}_{i}$ such as
the following sum-MSE is minimized
\begin{eqnarray}\label{MSEc}
\min\limits_{\mathbf{W}_{1},\mathbf{W}_{2},\mathbf{F}_{1},\mathbf{F}_{2},\mathbf{V}_{1},\mathbf{V}_{2}}~~\mathbf{MSE}_{sum}.
\end{eqnarray}
According to \eqref{R power}, we consider the following
transmission power constraint at relay node
\begin{eqnarray}\label{total R}
tr{\left(\textbf{F}_{1}\textbf{D}_{1}\textbf{F}^{H}_{1}+\textbf{F}_{2}\textbf{D}_{2}\textbf{F}^{H}_{2}\right)}\leq{P_{R_{1}}+P_{R_{2}}=P_{R}},
\end{eqnarray}
where
$\textbf{D}_{1}=\rho^{2}_{R}(\textbf{H}_{1,1}\mathbf{V}_{1}\mathbf{V}^{H}_{1}\textbf{H}^{H}_{1,1}+\textbf{H}_{1,2}\mathbf{V}_{2}\mathbf{V}^{H}_{2}\textbf{H}^{H}_{1,2}+\mathbf{I}_{N})$
and
$\textbf{D}_{2}=\rho^{2}_{R}(\textbf{H}_{2,1}\mathbf{V}_{1}\mathbf{V}^{H}_{1}\textbf{H}^{H}_{2,1}+\textbf{H}_{2,2}\mathbf{V}_{2}\mathbf{V}^{H}_{2}\textbf{H}^{H}_{2,2}+\mathbf{I}_{N})$.
The $P_{R_{i}}$ denotes the power constraint at the relay node
$\mathbf{R}_{i}$, and $P_{R}$ is the total relay power. The
transmission power constraint at two source nodes can be written
as
\begin{eqnarray}\label{source power}
tr{(\mathbf{V}_{i}\mathbf{V}^{H}_{i})}\leq{P_{i}},~~~~~ i=1,2
\end{eqnarray}
where $P_{i}$ is the available power at the $i$th source node.
According to \eqref{MSEc}, \eqref{total R} and \eqref{source
power}, the joint optimization problem of the sum-MSE can be
formulated as follows:
\begin{eqnarray}\label{MSEduality}
&\min\limits_{\mathbf{W}_{1},\mathbf{W}_{2},\mathbf{F}_{1},\mathbf{F}_{2},\mathbf{V}_{1},\mathbf{V}_{2}}&~~~~~~~~~~~~\mathbf{MSE}_{sum}\notag\\
&s.t.&~tr{\left(\textbf{F}_{1}\textbf{D}_{1}\textbf{F}^{H}_{1}+\textbf{F}_{2}\textbf{D}_{2}\textbf{F}^{H}_{2}\right)}\leq{P_{R}}\notag\\
&s.t.&~~~~~~~~tr{(\mathbf{V}_{i}\mathbf{V}^{H}_{i})}\leq{P_{i}}.
\end{eqnarray}
It is shown in \cite{b17} that at the optimum, $SNR_{1}=SNR_{2}$
holds true, thus leading to an SNR balancing result. Otherwise, if
$SNR_{1}>SNR_{2}$, then $P_{2}$ can be reduced to retain
$SNR_{1}=SNR_{2}$, and this reduction of $P_{2}$ will not violate
the power constraint, i.e.,
\begin{eqnarray}\label{MSEbalance}
P_{1}\cdot SNR_{1}=P_{2}\cdot SNR_{2}.
\end{eqnarray}
In Fig. \ref{2}, we show two examples of the SNR regions with
$\alpha_{1}=0.5$ and $\alpha_{2}=0.3$, where $\omega_{i}\in[0,1]$
is a Lagrange multiplier weight value and $\alpha_{i}\in[0,1]$ is
an SNR weight value. We have assumed the sum of SNR is a constant
value. It is clear that the SNR region of $\alpha_{1}$ is larger
than that of $\alpha_{2}$. For further details, see \cite{b17}.
As discussed in \cite{a5}, the optimization problems have the
performance matrix that are functions of SNR , namely the MSE at
the output of a linear-MMSE (LMMSE) filter of each user
\begin{eqnarray}\label{MSEbalance}
\mathbf{MSE}=\frac{1}{1+SNR}.
\end{eqnarray}
By these two approaches, the max-min optimization problem in
\eqref{MSEduality} can be efficiently written as
\begin{eqnarray}\label{MSEequal}
&\min\limits_{\mathbf{W}_{1},\mathbf{F}_{i},\mathbf{V}_{2}}&~~~~~~~~~~~~\mathbf{MSE}_{1}\\
&s.t.&~tr{(\textbf{F}_{1}\textbf{D}_{1}\textbf{F}^{H}_{1}+\textbf{F}_{2}\textbf{D}_{2}\textbf{F}^{H}_{2}})\leq{P_{R}}\label{MSE22}\\
&s.t.&~~~~~~\mathbf{MSE}_{1}=\mathbf{MSE}_{2}\label{MSE33},
\end{eqnarray}
where $i\in{1,2}$. Since the optimization problem \eqref{MSEequal}
is nonconvex, it is difficult to obtain the globally optimal
solution. In this paper, we present a locally optimal solution of
the joint optimization problem over $\mathbf{W}_{i}$,
$\mathbf{V}_{i}$ and $\mathbf{F}_{i}$ where $i=1,2$, which can be
solved by three stages, i.e., $1:$ The linear receiver weighted
matrices are optimized with the fixed source precoding matrix
$\mathbf{V}_{i}$ and relay amplifying matrices $\mathbf{F}_{i}$
($\mathbf{W}_{i}$ is not in constraints \eqref{MSE22} and
\eqref{MSE33}). $2:$ With given $\mathbf{W}_{i}$ and fixed
$\mathbf{F}_{i}$, update $\mathbf{V}_{i}$. $3:$ With given
$\mathbf{W}_{i}$ and $\mathbf{V}_{i}$, obtain suboptimal
$\mathbf{F}_{i}$ to solve \eqref{MSEequal}.

\emph{Lemma 1 }: For any fixed $\mathbf{V}_{i}$ and
$\mathbf{F}_{i}$, the minimization problems in \eqref{MSEequal}
are convex quadratic problems and the optimal $\mathbf{W}_{i}$ can
be obtained as the Wiener filter which is used to decode $s_{i}$
shown as follows
\begin{eqnarray}\label{MMSE}
\mathbf{W}^{o}_{i}=(\mathbf{\widetilde{H}}_{i}\mathbf{\widetilde{H}}^{H}_{i}+\mathbf{C}_{n_{i}})^{-1}\mathbf{\widetilde{H}}_{i},
\end{eqnarray}
\emph{Proof}: For source node $i$, the MSE can be further
expressed as:
\begin{eqnarray}\label{msep}
\mathbf{MSE}_{i}&=&\mathbf{W}^{H}_{i}\widetilde{\mathbf{H}}_{i}\widetilde{\mathbf{H}}^{H}_{i}\mathbf{W}_{i}-\mathbf{W}^{H}_{i}\widetilde{\mathbf{H}}_{i}-\widetilde{\mathbf{H}}^{H}_{i}\mathbf{W}_{i}\notag\\
&&+\mathbf{I}_{M}+\mathbf{W}^{H}_{i}\mathbf{C}_{n_{i}}\mathbf{W}_{i}
\end{eqnarray}
Based on \eqref{msep}, the derivation of an optimal MSE detection
matrix $\mathbf{W}^{opt}_{i}$ is equivalent to solving the
following equation:
\begin{eqnarray}\label{msep1}
\frac{\partial\mathbf{MSE}_{i}}{\partial\mathbf{W}^{H}_{i}}=2\widetilde{\mathbf{H}}_{i}\widetilde{\mathbf{H}}^{H}_{i}\mathbf{W}_{i}-2\widetilde{\mathbf{H}}_{i}+2\mathbf{C}_{n_{i}}\mathbf{W}_{i}=0.
\end{eqnarray}
Then, we may obtain closed-form solution of $\mathbf{W}_{i}$,
which is
\begin{eqnarray}\label{msep1}
\mathbf{W}^{o}_{i}=(\mathbf{\widetilde{H}}_{i}\mathbf{\widetilde{H}}^{H}_{i}+\mathbf{C}_{n_{i}})^{-1}\mathbf{\widetilde{H}}_{i}.
\end{eqnarray}
This completes the proof.

With the optimal $\mathbf{W}^{o}_{1}$ fixed, the outer
minimization problem in \eqref{MSEequal} can be rewritten as
\begin{eqnarray}\label{MSEopt}
&\min\limits_{\mathbf{F}_{1},\mathbf{F}_{2},\mathbf{V}_{1},\mathbf{V}_{2}}&~~~~~~~~~~~~\mathbf{MSE}^{o}_{1}\notag\\
&s.t.&~tr{(\textbf{F}_{1}\textbf{D}_{1}\textbf{F}^{H}_{1}+\textbf{F}_{2}\textbf{D}_{2}\textbf{F}^{H}_{2})}\leq {P_{R}}\notag\\
&s.t.&~~~~~~~\mathbf{MSE}_{1}=\mathbf{MSE}_{2},
\end{eqnarray}
where $\mathbf{MSE}^{o}_{1}$ is the MSE matrix using
$\mathbf{W}^{o}_{1}$. By substituting \eqref{MMSE} into
\eqref{mse}, we have
\begin{eqnarray}\label{MSEequ}
\mathbf{MSE}^{o}_{1}=[\mathbf{I}_{M}+\mathbf{\widetilde{H}}^{H}_{1}\mathbf{C}^{-1}_{n_{1}}\mathbf{\widetilde{H}}_{1}]^{-1}.~~~
i=1,2.
\end{eqnarray}
Note that the matrix inversion lemma is used to obtain
\eqref{MSEequ}.

\subsection{Joint Optimal Source and Relay Beamforming Matrices Design and Iterative Algorithm}

In this section, we focus on the source and relay beamforming
matrices design and develop an iterative algorithm which is
suboptimal for the general case, but has a much lower
computational complexity. For the fixed $\mathbf{F}_{i}$, the
source precoding matrix $\mathbf{V}_{i}$ is optimized by solving
the following problem
\begin{eqnarray}\label{MSEV}
&\min\limits_{\mathbf{V}_{1},\mathbf{V}_{2}}&~~~~~~~~~~~~tr\left[\mathbf{I}_{M}+\mathbf{V}^{H}_{2}\mathbf{\Phi}\mathbf{V}_{2}\right]^{-1}\notag\\
&s.t.&~~~tr\{\rho^{2}_{R}(\textbf{V}^{H}_{1}\mathbf{\Psi}_{1}\textbf{V}_{1}+\textbf{V}^{H}_{2}\mathbf{\Psi}_{2}\textbf{V}_{2})\}\leq{P_{R}}\notag\\
 &s.t.&~~~~~~~~~~~~~~~~tr\{\mathbf{V}^{H}_{i}\mathbf{V}_{i}\}\leq P_{i},
\end{eqnarray}
where
$\mathbf{\Phi}=\mathbf{\widehat{H}}^{H}_{1}\mathbf{\widehat{C}}^{-1}_{1}\mathbf{\widehat{H}}_{1}$
,
$\mathbf{\Psi}_{1}=\mathbf{H}^{H}_{1,1}\mathbf{F}^{H}_{1}\mathbf{F}_{1}\mathbf{H}_{1,1}+\mathbf{H}^{H}_{2,1}\mathbf{F}^{H}_{2}\mathbf{F}_{2}\mathbf{H}_{2,1}$,
and
$\mathbf{\Psi}_{2}=\mathbf{H}^{H}_{1,2}\mathbf{F}^{H}_{1}\mathbf{F}_{1}\mathbf{H}_{1,2}+\mathbf{H}^{H}_{2,2}\mathbf{F}^{H}_{2}\mathbf{F}_{2}\mathbf{H}_{2,2}$.
The Lagrangian function associated with the problem \eqref{MSEV}
is given by
\begin{eqnarray}\label{MSEL}
L_{\mathbf{V}}&=&tr\left[\mathbf{I}_{M}+\mathbf{V}^{H}_{2}\mathbf{\Phi}\mathbf{V}_{2}\right]^{-1}+\sum^{2}_{i=1}\mu_{i}\left(tr\{\mathbf{V}^{H}_{i}\mathbf{V}_{i}\}-P_{i}\right)\notag\\
&&+\mu_{3}\left\{tr\{\rho^{2}_{R}(\textbf{V}^{H}_{1}\mathbf{\Psi}_{1}\textbf{V}_{1}+\textbf{V}^{H}_{2}\mathbf{\Psi}_{2}\textbf{V}_{2})\}-P_{R}\right\},
\end{eqnarray}
where $\mu_{i}\geq0$ is the Lagrange multiplier.

\emph{Case 1}: When $\mu_{i}=0$, making the derivative of
$L_{\mathbf{V}}$ with respect to $\mathbf{V}_{2}$ be zero, we
obtain
\begin{eqnarray}\label{MSEL2}
\frac{\partial L_{\mathbf{V}}}{\partial
\mathbf{V}_{2}}=-\left[\mathbf{I}_{M}+\mathbf{V}^{H}_{2}\mathbf{\Phi}\mathbf{V}_{2}\right]^{-2}\mathbf{V}^{H}_{2}\mathbf{\Phi}=0.
\end{eqnarray}
Since $\mathbf{V}_{2}$ and $\mathbf{\Phi}$ are nonsingular
matrices, \eqref{MSEL2} can be represented as
\begin{eqnarray}\label{MSEL3}
\mathbf{I}_{M}+\mathbf{V}^{H}_{2}\mathbf{\Phi}\mathbf{V}_{2}=0.
\end{eqnarray}
Simplifying \eqref{MSEL3}, $\mathbf{I}_{M}>0$ and
$\mathbf{V}^{H}_{2}\mathbf{\Phi}\mathbf{V}_{2}\geq0$.
Consequently, in \emph{Case 1}, the optimal solution is not
existent.

\emph{Case 2}: When $\mu_{i}>0$, we rewrite the Lagrangian
function as
\begin{eqnarray}\label{MSEL4}
L_{\mathbf{V}}&=&\!\!\!\!\left[\mathbf{I}_{M}+\mathbf{V}^{H}_{2}\mathbf{\Phi}\mathbf{V}_{2}\right]^{-1}-\mu_{1}P_{1}-\mu_{2}P_{2}-\mu_{3}P_{R}\notag\\
&&\!\!\!\!+\mathbf{V}^{H}_{2}\Upsilon\Upsilon^{H}\mathbf{V}_{2}\!+\!\mu_{3}\rho^{2}_{R}\mathbf{V}_{1}^{H}\mathbf{\Psi}_{1}\mathbf{V}_{1}\!+\!\mu_{1}\mathbf{V}_{1}^{H}\mathbf{V}_{1}\!,
\end{eqnarray}
where
$\Upsilon\Upsilon^{H}=\mu_{2}\mathbf{I}_{M}+\mu_{3}\rho^{2}_{R}\mathbf{\Psi}_{2}$.
We obtain the derivative of $L_{\mathbf{V}}$ as
\begin{eqnarray}\label{MSEL5}
\frac{\partial L_{\mathbf{V}}}{\partial
\mathbf{V}_{2}}=-\left[\mathbf{I}_{M}+\mathbf{V}^{H}_{2}\mathbf{\Phi}\mathbf{V}_{2}\right]^{-2}\mathbf{V}^{H}_{2}\mathbf{\Phi}+\mathbf{V}^{H}_{2}\Upsilon\Upsilon^{H}=0.
\end{eqnarray}
Since $\mathbf{V}^{H}_{2}$ and $\mathbf{\Phi}$ are nonsingular
matrices, multiply both sides by
$\left(\mathbf{V}^{H}_{2}\right)^{-1}$ and $\mathbf{\Phi}^{-1}$,
we have
\begin{eqnarray}\label{MSEL6}
\left(\mathbf{V}^{H}_{2}\right)^{-1}\left[\mathbf{I}_{M}+\mathbf{V}^{H}_{2}\mathbf{\Phi}\mathbf{V}_{2}\right]^{-2}\mathbf{V}^{H}_{2}=\Upsilon\Upsilon^{H}\mathbf{\Phi}^{-1}.
\end{eqnarray}
Due to $\mathbf{\Phi}$ is Hermitian and positive definite, we
apply the Choleskey decomposition of
$\mathbf{\Phi}=\Omega^{H}\Omega$, where $\Omega$ is a lower
triangular matrix. Consequently, we represent \eqref{MSEL6} as
\begin{eqnarray}\label{MSEL7}
&&\left(\Omega^{H}\right)^{-1}\left(\mathbf{V}^{H}_{2}\right)^{-1}\left[\mathbf{I}_{M}+\mathbf{V}^{H}_{2}\Omega^{H}\Omega\mathbf{V}_{2}\right]^{-2}\mathbf{V}^{H}_{2}\Omega^{H}\notag\\
&=&\left(\Omega^{H}\right)^{-1}\Upsilon\Upsilon^{H}\left(\Omega^{H}\Omega\right)^{-1}\Omega^{H}.
\end{eqnarray}
By the definition of the matrix identity as
\begin{eqnarray}\label{ID}
\left[\mathbf{I}_{M}+\mathbf{X}\mathbf{X}^{H}\right]^{-1}\mathbf{X}=\mathbf{X}\left[\mathbf{I}_{N}+\mathbf{X}^{H}\mathbf{X}\right]^{-1},
\end{eqnarray}
for any $M\times N$ matrix $\mathbf{X}$, we can rewrite
\eqref{MSEL7} as
\begin{eqnarray}\label{MSEL8}
\left[\mathbf{I}_{M}+\Omega\mathbf{V}_{2}\mathbf{V}^{H}_{2}\Omega^{H}\right]^{-2}=\left(\Omega^{H}\right)^{-1}\Upsilon\Upsilon^{H}\Omega^{-1}.
\end{eqnarray}
Solving \eqref{MSEL8} for $\mathbf{V}_{2}$, we obtain
\eqref{MSEL7} as
\begin{eqnarray}\label{MSEL8}
\mathbf{V}_{2}=\left(\nabla\nabla^{H}-\mathbf{\Phi}^{-1}\right)^{\frac{1}{2}},
\end{eqnarray}
where $\nabla=\left(\Upsilon^{H}\Omega\right)^{-\frac{1}{2}}$.
Obviously, the precoding matrix ${\textbf{V}}_1$ can be obtained
in the same way.

\begin{figure}
\begin{center}
\includegraphics [width=70mm,height=30mm]{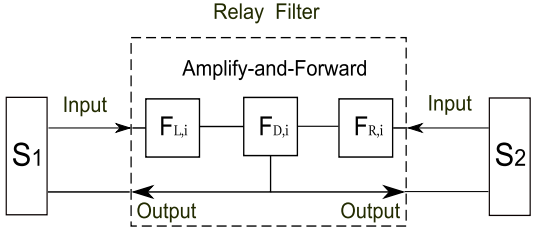}
\caption{\label{3} The relay filter design of the proposed QL-QR
technique.}
\end{center}
\end{figure}

Figure 3 shows our proposed relay filter design, which forwards
the received signal (input) from $\mathbf{S}_{1}$ amplified by a
Left Filter (LF) matrix $\mathbf{F}_{L,i}$ and the signal from
$\mathbf{S}_{2}$ amplified by a Right Filter (RF) matrix
$\mathbf{F}_{R,i}$ to the Center Filter (CF) $\mathbf{F}_{D,i}$
that amplifies the outputs from the Left Filter (LF) matrix
$\mathbf{F}_{L,i}$ and the Right Filter (RF) matrix
$\mathbf{F}_{R,i}$, and forward them to $\mathbf{S}_{1}$ and
$\mathbf{S}_{2}$ (output).\footnote{For example: For
$\mathbf{S}_{1}$, the equivalent channel can be written as
$\widetilde{\mathbf{H}}_{1}=\textbf{H}^{T}_{1,1}\textbf{F}_{1}\textbf{H}_{1,2}+\textbf{H}^{T}_{2,1}\textbf{F}_{2}\textbf{H}_{2,2}=\textbf{H}^{T}_{1,1}\textbf{F}_{L,1}\textbf{F}_{D,1}\textbf{F}_{R,1}\textbf{H}_{1,2}+\textbf{H}^{T}_{2,1}\textbf{F}_{L,2}\textbf{F}_{D,2}\textbf{F}_{R,2}\textbf{H}_{2,2}$.
For $\mathbf{S}_{2}$, the equivalent channel can be written as
$\widetilde{\mathbf{H}}_{2}=\textbf{H}^{T}_{1,2}\textbf{F}_{1}\textbf{H}_{1,1}+\textbf{H}^{T}_{2,2}\textbf{F}_{2}\textbf{H}_{2,1}=\textbf{H}^{T}_{1,2}\textbf{F}_{R,1}\textbf{F}_{D,1}\textbf{F}_{L,1}\textbf{H}_{1,1}+\textbf{H}^{T}_{2,1}\textbf{F}_{R,2}\textbf{F}_{D,2}\textbf{F}_{L,2}\textbf{H}_{2,2}$.}

\emph{Lemma 2}: The optimal relay filter constructive of
$\mathbf{F}_{L}$, $\mathbf{F}_{R}$ and $\mathbf{F}_{D}$ matrices
i.e., for $\mathbf{R}_{1}$ and $\mathbf{R}_{2}$ can be designed as
\begin{eqnarray}\label{RL+RF}
&&\mathbf{F}_{L,1}=\mathbf{Q}^{*}_{L,1},~\mathbf{F}_{L,2}=\mathbf{Q}^{*}_{L,2},\notag\\
&&\mathbf{F}_{R,1}=\mathbf{Q}^{H}_{R,1},~\mathbf{F}_{R,2}=\mathbf{Q}^{H}_{R,2},\notag\\
&&\mathbf{F}_{D,1}~\mathrm{and}~ \mathbf{F}_{D,2}~\mathrm{ are~
diagonal~ matrix}.
\end{eqnarray}
\emph{Proof}: For $\mathbf{F}_{L,i}$ and $\mathbf{F}_{R,i}$, the
proof is similar as \emph{Theorem 3.1} in \cite{a6}. For
$\mathbf{F}_{D,i}$, using \emph{Theorem 2} in \cite{b8}, the
structure of $\mathbf{F}_{D,i}$ is optimal for the two cases of
\begin{eqnarray}\label{OP1}
&&(a):~~~~~R(\mathbf{H}_{1}\mathbf{V}_{1})\bot R(\mathbf{H}_{2}\mathbf{V}_{2}); (\mathbf{H}^{*}_{1})\bot R(\mathbf{H}^{*}_{2})\notag\\
&& ~~\mathrm{and}~\, ~~~R(\mathbf{H}_{3}\mathbf{V}_{1})\bot R(\mathbf{H}_{4}\mathbf{V}_{2}); (\mathbf{H}^{*}_{3})\bot R(\mathbf{H}^{*}_{4})\notag\\
 &&(b):~~~~~R(\mathbf{H}_{1}\mathbf{V}_{1})\| R(\mathbf{H}_{2}\mathbf{V}_{2}); (\mathbf{H}^{*}_{1})\bot R(\mathbf{H}^{*}_{2})\notag\\
&& ~~\mathrm{and}~\, ~~~R(\mathbf{H}_{1}\mathbf{V}_{1})\|
R(\mathbf{H}_{2}\mathbf{V}_{2}); (\mathbf{H}^{*}_{1})\bot
R(\mathbf{H}^{*}_{2}).
\end{eqnarray}
\emph{Case a}: If $N=2$, the optimal $\mathbf{F}_{D,i}$ is a
diagonal matrix given as
\begin{eqnarray}\label{OP4}
\mathbf{F}_{D,i} &\triangleq& {\left[\begin{array}{cc}
              \mathbf{f}_{d,i,1} & \mathbf{0} \\
             \mathbf{0} & \mathbf{f}_{d,i,2} \\
             \end{array}\right]}.
             \end{eqnarray}
If $N=2a$, $a={2,3,...}$, the optimal $\mathbf{F}_{D,i}$ is a
$2\times2$ block diagonal matrix given as
\begin{eqnarray}\label{OP2}
\mathbf{F}_{D,i} &\triangleq& {\left[\begin{array}{cc}
              \mathbf{F}_{D,i,1} & \mathbf{0} \\
             \mathbf{0} & \mathbf{F}_{D,i,2} \\
             \end{array}\right]},
             \end{eqnarray}
where $\mathbf{F}_{D,i,1}$ and $\mathbf{F}_{D,i,2}$ are
$\frac{N}{2}\times \frac{N}{2}$ matrices. \\
\emph{Case b}: The optimal $\mathbf{F}_{D,i}$ is defined as
\begin{eqnarray}\label{OP6}
\mathbf{F}^{\star}_{D,i} &\triangleq& {\left[\begin{array}{cc}
              \mathbf{f}_{d,i,1} \\
             \mathbf{f}_{d,i,2} \\
             \end{array}\right]}.
             \end{eqnarray}
\emph{Discussion}: In \emph{Case b}, since
$\mathbf{F}^{\star}_{D,i}$ is optimal, but the computational
complexity will be considerably increased compared with \emph{Case
a}, so we exclude it.

For \emph{Case a}, Before we develop a numerical method to solve
vector $\mathbf{F}_{D,i}$, let us have some insights into the
structure of this suboptimal relay beamforming matrix. To simplify
relay beamforming matrix $\mathbf{F}_{D,i}$, we introduce a
following property:

\emph{Property 1}: The statistical behavior of a unitary matrix
$\mathbf{U}$ remains unchanged when multiplied by any unitary
matrix $\mathbf{T}$ independent of $\mathbf{U}$. In other worlds,
$\mathbf{TU}$ has the same distribution as $\mathbf{U}$, i.e., in
\eqref{RL+RF},
\begin{eqnarray}\label{Prop}
&&|\mathbf{F}_{1}|=|\mathbf{F}_{L,1}\mathbf{F}_{D,1}\mathbf{F}_{R,1}|=|\mathbf{F}_{D,1}|\notag\\
&&|\mathbf{F}_{2}|=|\mathbf{F}_{L,2}\mathbf{F}_{D,2}\mathbf{F}_{R,2}|=|\mathbf{F}_{D,2}|.
\end{eqnarray}
Now, let us introduce the
following $\mathbf{QL}$ decompositions
\begin{eqnarray}\label{LQ}
\left[\mathbf{H}_{1,1}\mathbf{V}_{1},
\mathbf{H}_{2,1}\mathbf{V}_{1}\right]&=&\left[\mathbf{Q}_{L,1}\mathbf{L}_{1},
\mathbf{Q}_{L,2}\mathbf{L}_{2} \right],
\end{eqnarray}
where $\mathbf{Q}_{L,i}$ for $i=1,2$, is a unitary matrix with a
dimension ${\mathbb{C}^{N\times{N}}}$,
and $\left\{\mathbf{L}_{1}, \mathbf{L}_{2}\right\}\in
{\mathbb{C}^{N\times{M}}}$ are lower triangular matrices. 
Similarly, let us introduce another decomposition, namely,
$\mathbf{QR}$ decomposition as
\begin{eqnarray}\label{QR}
\left[\mathbf{H}_{1,2}\mathbf{V}_{2},
\mathbf{H}_{2,2}\mathbf{V}_{2}\right]&=&\left[\mathbf{Q}_{R,1}\mathbf{R}_{1},
\mathbf{Q}_{R,2}\mathbf{R}_{2} \right],
\end{eqnarray}
where $\mathbf{Q}_{R,i}\in{\mathbb{C}^{N\times{M}}}$ for $i=1,2$,
is a unitary matrix, and $\left\{\mathbf{R}_{1},
\mathbf{R}_{2}\right\} \in{\mathbb{C}^{M\times{M}}}$ are upper
triangular matrices. Substituting \eqref{LQ}, \eqref{QR}, and
\eqref{RL+RF} back into \eqref{s3}, i.e,. for $\mathbf{S}_{1}$, we
may get equivalent received signals shown as,
\begin{eqnarray}\label{equ}
\widehat{y}_{1}&=&\!\!\!\!(\mathbf{L}^{T}_{1}\mathbf{F}_{D,1}\mathbf{R}_{1}+\mathbf{L}^{T}_{2}\mathbf{F}_{D,2}\mathbf{R}_{2})x_{2}\notag\\
&&\!\!\!\!+\mathbf{L}^{T}_{1}\mathbf{F}_{D,1}n_{R_{1}}\!+\!\mathbf{L}^{T}_{2}\mathbf{F}_{D,2}n_{R_{2}}\!+\!n_{2}\notag\\
&=&\!\!\!\!\mathbf{\widehat{H}}_{1,2}x_{2}+\widehat{n}_{1},
\end{eqnarray}
where
$\mathbf{\widehat{H}}_{1}=\mathbf{L}^{T}_{1}\mathbf{F}_{D,1}\mathbf{R}_{1}+\mathbf{L}^{T}_{2}\mathbf{F}_{D,2}\mathbf{R}_{2}$
and
$\widehat{n}_{1}=\mathbf{L}^{T}_{1}\mathbf{F}_{D,1}n_{R_{1}}\!+\!\mathbf{L}^{T}_{2}\mathbf{F}_{D,2}n_{R_{2}}\!+\!n_{2}$
are efficient channel and noise coefficients, obtained from the
covariance of $\widehat{n}_{1}$, we have
\begin{eqnarray}\label{equa}
\mathbf{\widehat{C}}_{1}&=&\widehat{n}_{1}\widehat{n}^{H}_{1}\notag\\
&=&\mathbf{L}^{T}_{1}\mathbf{F}_{D,1}\mathbf{F}^{H}_{D,1}\mathbf{L}^{*}_{1}+\mathbf{L}^{T}_{2}\mathbf{F}_{D,2}\mathbf{F}^{H}_{D,2}\mathbf{L}^{*}_{2}+\mathbf{I}_{N}.
\end{eqnarray}

For fixed $\mathbf{V}_{1}$ and $\mathbf{V}_{2}$, using \eqref{equ}
and \emph{Property 1}, the optimal problem \eqref{MSEopt} becomes
\begin{eqnarray}\label{MSEnew}
&\max\limits_{\mathbf{F}_{D,1},
\mathbf{F}_{D,2}}&~~~tr\left(\mathbf{I}_{N}+\mathbf{\widehat{H}}^{H}_{1}\mathbf{\widehat{C}}_{1}^{-1}\mathbf{\widehat{H}}_{1}\right)\label{MSEnew}\\
&s.t.&tr{(\textbf{F}^{H}_{D,1}\mathbf{D}_{1}\textbf{F}_{D,1}\!+\!\textbf{F}^{H}_{D,2}\mathbf{D}_{2}\textbf{F}_{D,2})}\!\leq\!{P_{R}}.
\end{eqnarray}
Then, \eqref{MSEnew} can be represented as
\begin{eqnarray}\label{MSEnew1}
tr\left((\mathbf{\widehat{H}}^{H}_{1}\mathbf{\widehat{C}}_{1}^{-1}\mathbf{\widehat{H}}_{1})+n\right),
\end{eqnarray}
where the lemma $tr(\mathbf{A+B})=tr(\mathbf{A})+tr(\mathbf{B})$
has been used. Since the matrix $\mathbf{\widehat{C}}_{1}$ is
Hermitian and positive definite, we can decompose this matrix
using Cholesky factorization as
\begin{eqnarray}\label{Choleskey}
\mathbf{\widehat{C}}_{1}&=&\mathbf{\Xi}^{H}_{1}\mathbf{\Xi}_{1}
\end{eqnarray}
where $\mathbf{\Xi}_{1}$ denote a lower triangular matrix.
By substituting \eqref{Choleskey} back into \eqref{MSEnew1}, we
can simply rewrite the optimal problem as
\begin{eqnarray}\label{MSEnew2}
&&\mathrm{max}~\left(\mathbf{MSE}^{o}_{1}\right)^{-1}\!\!\notag\\
&=&\mathrm{max}~tr\left((\mathbf{\widehat{H}}^{H}_{1}\left(\mathbf{\Xi}^{H}_{1}\mathbf{\Xi}_{1}\right)^{-1}\mathbf{\widehat{H}}_{1})+n\right)\notag\\
&\overset{a}{=}&\mathrm{max}~tr\left(\left(\mathbf{\widehat{H}}^{H}_{1}\mathbf{\Xi}^{-1}_{1}\right)\left(\mathbf{\widehat{H}}^{H}_{1}\mathbf{\Xi}^{-1}_{1}\right)^{H}\right)\notag\\
&=&\mathrm{max}~tr\left(\mathbf{B}_{1}\mathbf{B}^{H}_{1}\right),
\end{eqnarray}
where $\overset{a}{=}$ denotes $n$ has nothing to do with the
maximum solution and
$\mathbf{B}_{1}=\mathbf{\widehat{H}}^{H}_{1}\mathbf{\Xi}^{-1}_{1}$.
Thus, the optimal problem can be represented as the determinant
maximization of $|\mathbf{B}_{1}|^{2}$.

In \emph{Case a}, since $\mathbf{F}_{D,i}$ is the block diagonal
matrix, its determinant can be written as
\begin{eqnarray}\label{target}
\mathbf{det}{\mathbf{F}_{D,i}}=\mathbf{det}{\mathbf{F}_{D,i,1}}\cdot\mathbf{det}{\mathbf{F}_{D,i,2}}.
\end{eqnarray}
Let $\mathbf{A}$, $\mathbf{B}$, $\mathbf{C}$, and $\mathbf{D}$ be
an $\frac{N}{4}\times\frac{N}{4}$ matrix. We can define
$\mathbf{det}{\mathbf{F}_{D,i,i}}$, for $i\in{1,2}$ as,
\begin{eqnarray}\label{target1}
\mathbf{det}{\mathbf{F}_{D,i,i}} &=& {\left|\begin{array}{cc}
              \mathbf{A} & \mathbf{D} \\
             \mathbf{B} & \mathbf{C} \\
             \end{array}\right|}\notag\\
&=& {\left|\begin{array}{cc}
              \mathbf{A} & \mathbf{0} \\
             \mathbf{B} & \mathbf{I} \\
             \end{array}\right|}
             {\left|\begin{array}{cc}
              \mathbf{I} & \mathbf{A}^{-1} \\
             \mathbf{0} & \mathbf{C}-\mathbf{BA^{-1}D} \\
             \end{array}\right|}\notag\\
&=& |\mathbf{A}|\left|\mathbf{C}-\mathbf{BA^{-1}D}\right|,
             \end{eqnarray}
where $\mathbf{I}$ stands for an $\frac{N}{4}\times\frac{N}{4}$
identity matrix. In \eqref{target1}, to obtain maximum
$\mathbf{det}{\mathbf{F}_{D,i,i}}$, we should minimize
$\mathbf{BA^{-1}D}$. Let us introduce the SVD of $\mathbf{B}$,
$\mathbf{A}$, and $\mathbf{D}$ as
\begin{eqnarray}\label{target3}
\mathbf{B}=\mathbf{U}_{B}\mathbf{\Sigma}_{B}\mathbf{\Lambda}^{H}_{B},~
\mathbf{A}=\mathbf{U}_{A}\mathbf{\Sigma}_{A}\mathbf{\Lambda}^{H}_{A},~
\mathbf{D}=\mathbf{U}_{D}\mathbf{\Sigma}_{D}\mathbf{\Lambda}^{H}_{D},
\end{eqnarray}
where $\mathbf{U}_{i}$, $\mathbf{\Lambda}_{i}$, $i\in{\left\{A, B,
D\right\}}$, are the unitary matrices, and $\mathbf{\Sigma}_i$ is
an $\frac{N}{4}\times\frac{N}{4}$ diagonal matrix. Substituting
\eqref{target3} back into $\mathbf{BA^{-1}D}$, we have
\begin{eqnarray}\label{target4}
&&\min~tr\left(\mathbf{U}_{B}\mathbf{\Sigma}_{B}\mathbf{\Lambda}^{H}_{B}\left(\mathbf{U}_{A}\mathbf{\Sigma}_{A}\mathbf{\Lambda}^{H}_{A}\right)^{-1}\mathbf{U}_{D}\mathbf{\Sigma}_{D}\mathbf{\Lambda}^{H}_{D}\right)\notag\\
&\triangleq&\min~tr\left(\mathbf{\Sigma}_{B}\left(\mathbf{\Sigma}_{A}\right)^{-1}\mathbf{\Sigma}_{D}\right)\notag\\
&\triangleq&\min_{b_{i},d_{i}}~\sum^{n/4}_{i}\frac{b_{i}d_{i}}{a_{i}},
\end{eqnarray}
where $b_{i}$, $d_{i}$, and $a_{i}$ are the diagonal elements of
$\mathbf{\Sigma}_{B}$, $\mathbf{\Sigma}_{D}$ and
$\mathbf{\Sigma}_{A}$, respectively. To simplify our discussion,
we assume $\mathbf{F}_{D,i,i}$ is a semi-positive matrix, thus, we
have the minimum solution as $b_{i}d_{i}=0$. Interestingly, if
both $b_{i}$ and $d_{i}$ are $0$, $\mathbf{F}_{D,i,i}$ is a
diagonal matrix. Otherwise, it is a lower/upper triangular matrix.
In addition, for $\mathbf{S}_{1}$, the equivalent channel
$\widehat{\mathbf{H}}_{1}$, since the terms $\mathbf{L}^{T}_{1}$,
$\mathbf{L}^{T}_{2}$, $\mathbf{R}_{1}$, and $\mathbf{R}_{2}$ are
upper triangular matrices, the optimal $\mathbf{F}_{D,i}$ should
be an upper triangular matrix. Since the equivalent channel
$\widehat{\mathbf{H}}_{2}$, $\mathbf{L}_{1}$, $\mathbf{L}_{2}$,
$\mathbf{R}^{T}_{1}$, and $\mathbf{R}^{T}_{2}$ are lower
triangular matrices for $\mathbf{S}_{2}$, the optimal
$\mathbf{F}_{D,i}$ is a lower triangular matrix. Therefore, if and
only if $\mathbf{F}_{D,i}$ is a diagonal matrix,
the sum-MSE is optimal in our proposed method. 
This completes the proof for \emph{Lemma 2}.

\emph{Property 2}: For any $M\times N$ rectangular matrices
$\mathbf{G}$ and $\mathbf{J}$, matrices $\mathbf{A}$ and
$\mathbf{B}$ are lower/upper triangular matrices based on QR or QL
decomposition of $\mathbf{G}$ and $\mathbf{J}$. If
$a_{i,i}+b_{i,i}\neq0$, where $a_{i,i}$ and $b_{i,i}$ are diagonal
elements of matrices $\mathbf{A}$ and $\mathbf{B}$, respectively,
we can easily obtain
\begin{eqnarray}\label{det}
\mathrm{det}\left(\mathbf{A}+\mathbf{B}\right)=\prod^{m}_{i=1}(a_{i,i}+b_{i,i})
\geq\mathrm{det}\mathbf{A}+\mathrm{det}\mathbf{B},
\end{eqnarray}
Consequently, we have
\begin{eqnarray}\label{det1}
\mathrm{det}\mathbf{\widehat{H}}^{H}_{1}&=&\prod^{m}_{i=1}\left(l_{1,i,i}f_{D,1,i}r_{1,i,i}+l_{2,i,i}f_{D,2,i}r_{2,i,i}\right)\notag\\
&=&\prod^{m}_{i=1}(\varsigma_{1}+\varsigma_{2}),
\end{eqnarray}
where $\varsigma_{1}=l_{1,i,i}f_{D,1,i}r_{1,i,i}$,
$\varsigma_{2}=l_{2,i,i}f_{D,2,i}r_{2,i,i}$, $l_{1,i,i}$,
$f_{D,1,i}$, $r_{1,i,i}$, $l_{2,i,i}$, $f_{D,2,i}$, and
$r_{2,i,i}$ are diagonal elements of $\mathbf{L}_{1}$,
$\mathbf{F}_{D,1}$, $\mathbf{R}_{1}$, $\mathbf{L}_{2}$,
$\mathbf{F}_{D,2}$, and $\mathbf{R}_{2}$, respectively. Since
$\mathbf{\Xi}_{1}$ and $\mathbf{\Xi}_{1}^{-1}$ are also lower
triangular matrices, we have
\begin{eqnarray}\label{det2}
\mathrm{det}\mathbf{B}_{1}&=&\mathrm{det}\mathbf{\widehat{H}}^{H}_{1}\mathrm{det}\mathbf{\Xi}_{1}^{-1}\notag\\
&=&\prod^{m}_{i=1}(\varsigma_{1}+\varsigma_{2})\xi_{i}
\end{eqnarray}
where $\xi_{i}$ is the diagonal element of
$\mathbf{\Xi}^{-1}_{1}$.
%
Now, the optimization problem can be reformulated as
\begin{eqnarray}
&\max\limits_{\mathbf{F}_{D,1},\mathbf{F}_{D,2}}&~~~~~~~~~~~~~~~~~~~~|\mathbf{B}_{1}|^{2}\label{optnew}\\
&s.t.&{\left(\!\!\!\begin{array}{cc}
              |\mathbf{B}_{1}|^{2} & \mathbf{I}_{N} \\
              \mathbf{I}_{N} & \!\!\!\mathbf{I}_{N}+\mathbf{\Xi}^{-1}_{1}\widehat{\mathbf{H}}^{H}_{1}\widehat{\mathbf{H}}_{1}\left(\mathbf{\Xi}^{-1}_{1}\right)^{H} \\
              \end{array}\!\!\!\!\!\right)}\label{optnew1}\\
&s.t.&tr{(\textbf{F}_{D,1}\mathbf{D}_{1}}\textbf{F}^{H}_{D,1}+\textbf{F}_{D,2}\mathbf{D}_{2}\textbf{F}^{H}_{D,2})\!\leq\!{P_{R}}\label{optnew2},
\end{eqnarray}
It is clear that \eqref{optnew}-\eqref{optnew2} is a convex
problem for beamformer vectors $\mathbf{F}_{D,1}$ and
$\mathbf{F}_{D,2}$, which can be efficiently solved by the
interior-point method \cite{a7}.

In summary, we outline the iterative beamforming design algorithm
as follows ($\mathbf{QL}-\mathbf{QR}~\mathbf{Algorithm}$):

\alglanguage{pseudocode}
\begin{algorithm}[h]
\caption{QL-QR Algorithm} $\mathbf{1}$.~~$\mathbf{Initialize}$:
$\mathbf{F}^{(n)}_{i}$, $\mathbf{W}^{(n)}_{i}$,
$\mathbf{V}^{(n)}_{i}$, $\mathbf{H}^{(n)}_{i,j}$,
for $i,j=1,2$, set $n=0$;\\
$\mathbf{2}$.~~$\mathbf{Repeat}$:
\begin{algorithmic}[1]
\For {$n\leftarrow n+1$}\\
for given $\mathbf{F}^{(n-1)}_{i}$, $\mathbf{H}^{(n-1)}_{i,j}$ and
$\mathbf{V}^{(n-1)}_{i}$ update $\mathbf{W}^{(n)}_{i}$ using
(22);\\
for fixed $\mathbf{F}^{(n-1)}_{i}$, $\mathbf{H}^{(n-1)}_{i,j}$,
$\mathbf{W}^{(n)}_{i}$ update $\mathbf{V}^{(n)}_{i}$ using
(35);\\
decompose $(\mathbf{H}^{(n)}_{1,1}\mathbf{V}^{(n)}_{1},
\mathbf{H}^{(n)}_{2,1}\mathbf{V}^{(n)}_{1})$ using $\mathbf{QL}$
decomposition as (42);\\
decompose $(\mathbf{H}^{(n)}_{1,2}\mathbf{V}_{2}^{(n)},
\mathbf{H}^{(n)}_{2,2}\mathbf{V}^{(n)}_{2})$ using $\mathbf{QR}$
decomposition as
(43);\\
compute
$\widehat{\mathbf{C}}^{(n)}_{1}$;\\
decompose $\widehat{\mathbf{C}}^{(n)}_{1}$ using Cholesky decomposition as (49);\\
for fixed $\mathbf{\widehat{H}}_{1}$, compute $\mathrm{det}\mathbf{B}^{(n)}_{1}$ using (57);\\
$\mathbf{Until}$ $\mathbf{MSE}_{1}$ converges.
      \EndFor
\end{algorithmic}
\end{algorithm}

Since in the QL-QR algorithm, the solution of each subproblem is
optimal, we conclude that the total MSE value is decreased as the
number of iterations increases. Meanwhile, the total MSE is lower
bounded.

\begin{figure}
\begin{center}
\includegraphics [width=70mm,height=24mm]{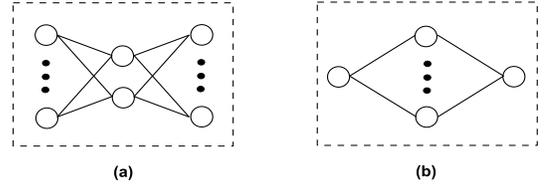}
\caption{\label{4} The extended system model. (a): The $K$ pair
source nodes scenario. (b): The $T$ relay nodes scenario.}
\end{center}
\end{figure}

\emph{Discussion}: The extended two system models are shown in
Fig. \ref{4}, which are the multipair scenario with two relay
nodes and $K$ pair source nodes, and the $Z$ ($Z$ should be even
number) relay nodes scenario with two source nodes. In Figure
4(a), each pair sources and two relay nodes can be seen as a
group. Since each pair source nodes are independent with each
other, we can design that there are $K$ RFs , LFs and one CF
equipped at each relay nodes. Therefor, the extended system model
(a) can be seen as $K$ parallel of our proposed system model. In
Figure 4(b), two source nodes and every two relay nodes can be
seen as one group. Obviously, the extended system model (b) can be
seen as $\frac{Z}{2}$ parallel of our proposed system model.

\section{Computational Complexity Analysis}
In this section, we measure the performance of the proposed QL-QR
scheme in terms of the computational complexity compared with
existing algorithms by using the total number of floating point
operations (FLOPs). A flop is defined as a real floating
operation, i.e., a real addition, multiplication, division, and so
on. In \cite{c1}, the authors show the computational complexity of
the real Choleskey decomposition. For complex numbers, a
multiplication followed by an addition needs 8 FLOPs, which leads
to 4 times its real computation. According to \cite{a3}, the
required number of FLOPs of each matrix is described as follows:\\\\
1. Multiplication of $m\times{n}$ and $n\times{p}$ complex
matrices: $8mnp-2mp$;\\
2. Multiplication of $m\times{n}$ and $n\times{m}$ complex
matrices: $4nm\times{(m+1)}$;\\
3. SVD of an $m\times{n}(m\leq{n})$ complex matrix where only
$\Sigma$ is obtained: $32(mn^{2}-n^{3}/3)$;\\
4. SVD of an $m\times{n}(m\leq{n})$ complex matrix where only
$\Sigma$ and $\Lambda$ are obtained: $32(nm^{2}+2m^{3})$,\\
5. SVD of an $m\times{n}(m\leq{n})$ complex matrix where
$U$,$\Sigma$, and $\Lambda$ are obtained:
$8(4n^{2}m+8nm^{2}+9m^{3})$;\\
6. Inversion of an $m\times{m}$ real matrix using Gauss-Jordan
elimination: $2m^{3}-2m^{2}+m$;\\
7. Cholesky factorization of an $m\times{m}$ complex matrix:
$8m^{3}/3$.\\
8. QR or QL decomposition of an $m\times n$ conplex matrix
$16\left(n^{2}m-nm^{2}+\frac{1}{3}m^{3}\right)$.\\

For the conventional RBD method \cite{b18}, the authors consider a
linear MU-MIMO precoding scheme for DL MIMO systems. For the
non-regenerative MIMO relay systems \cite{b19}, the authors
investigate a precoding design for a 3-node MIMO relay network. In
\cite{c2}, a relay-aided system based on a quasi-EVD channel is
proposed. We compare the required number of FOLPs of our proposed
method with conventional precoding algorithm, such as the
conventional RBD, the non-regenerative MIMO relay system, and the
CD-BD algorithm as shown in Tables I, II, III, and IV,
respectively, under the assumption that $N_{T}=N_{R}$ and
$\overline{N}_{i}=N_{T}-N_{i}$.

\begin{figure}
\begin{center}
\includegraphics [width=92mm,height=75mm]{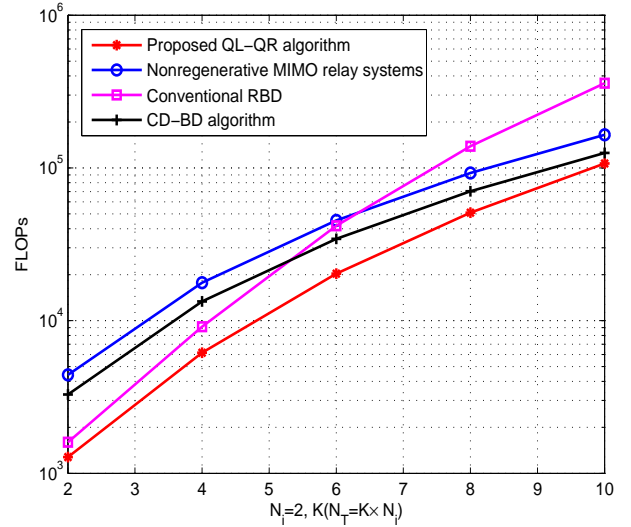}\\
\caption{\label{5} The complexity comparisons for required FLOPs
versus the number of the users $K$.}
\end{center}
\end{figure}

\begin{figure}
\begin{center}
\includegraphics [width=92mm,height=75mm]{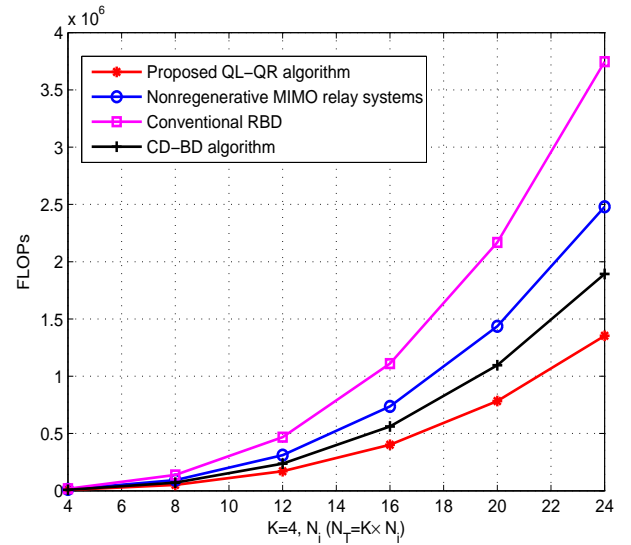}\\
\caption{\label{6} The complexity comparisons for required FLOPs
versus the number of the receive antennas $N_{i}$ for each user.}
\end{center}
\end{figure}

For instance, the $(2,2,2)\times6$ case denotes a system with
three users $(K=3)$, where each user is equipped with two antennas
$(N_{i}=2)$ and the total number of transmit antennas is six
$(N_{T}=2\times3=6)$. The required number FLOPs of the QL-QR
algorithm, the conventional RBD, the non-regenerative MIMO relay
system, and the CD-BD algorithm are counted as 33530, 40824,
45306, 34638, respectively. From these results, we can see that
the reduction in the number of FLOPs of our proposed precoding
method is $17.87\%$, $25.99\%$, and $3.20\%$ on an individual
basis compared to the conventional RBD, the non-regenerative MIMO
relay systems, and the CD-BD algorithm. Thus, our proposed QL-QR
algorithm exhibits lower complexity than conventional algorithms.
In addition, the complexity reduces as $N_{i}$ and $N_{T}$
increase with fixed $K$.

We summarize our calculation results of the required number of
FLOPs of the alternative methods in Tables I, II, III, and IV and
show them in Figures \ref{5} and \ref{6}. Figure \ref{5} shows the
computational complexity where $N_{i}=2$ and a value of $K$
varies. And Figure \ref{6} shows the computational complexity
where $K=4$ and a value of $N_{i}$ varies. For the conventional
RBD method, the orthogonal complementary vector $\mathbf{V}_{k,0}$
requires $K$ times SVD operations. If only $\mathbf{V}_{k,0}$ is
obtained, it is not computationally efficient. In step 5, the
efficient channel
$\mathbf{H}_{eff}=\mathbf{H}_{i}\mathbf{P}^{a}_{i}$ is decomposed
by the SVD with a dimension $\mathbf{R}_{eff}\times N_{T}$, where
$\mathbf{R}_{eff}$ is the rank of $\mathbf{H}_{eff}$. In the
nonregenerative MIMO relay method and the CD-BD algorithm, two SVD
operations are performed for the channels from the source to relay
and from relay to the destination, and then the efficient channel
covariance matrix is measured. In the nonregenerative MIMO relay
method, the authors compute $\mathbf{A}$ using the EVD an then
they diagonalize $\mathbf{G}$. In the CD-BD algorithm, the authors
calculate $\mathbf{V}^{a}_{i}$ by the SVD of
$\mathbf{H}^{\dag}_{mse}$ and then they structure
$\mathbf{V}^{b}_{i}$ by using the Choleskey decomposition.

In our proposed QL-QR algorithm, we take advantage of QL and QR
decompositions instead of the SVD operation, and then we compute
an efficient channel as well as decompose a noise covariance
matrix by the Choleskey decomposition. Finally, we calculate the
determinant of $\mathbf{B}^{2}_{i}$ to solve an optimization
problem. Obviously, our proposed QL-QR algorithm outperforms
conventional algorithms in the light of the computational
complexity.

\begin{table*}[tbp]
\centering  
\caption{Computational complexity of the proposed QL-QR
Algorithm.}
\begin{tabular}{| c | c | c | c |} \hline 
Step   & Operations  & FLOPS & Case: $(2,2,2)\times 6$ \\ \hline  
1 &$\mathbf{V}_{1}, \mathbf{V}_{2}$ &$2\times
K(40N^{3}_{i}-24N^{2}_{i}+17N_{i})$ & 1560
\\ \hline        
2 &$\mathbf{Q}_{L,1}\mathbf{L}_{1},
\mathbf{Q}_{L,2}\mathbf{L}_{2}$
&$2\times16K(N^{2}_{T}N_{i}-N_{T}N^{2}_{i}+\frac{1}{3}N^{3}_{i})$
& 4864
\\ \hline        
3 &$\mathbf{Q}_{R,1}\mathbf{R}_{1},
\mathbf{Q}_{R,2}\mathbf{R}_{2}$
&$2\times16K(N^{2}_{T}N_{i}-N_{T}N^{2}_{i}+\frac{1}{3}N^{3}_{i})$
& 4864
\\ \hline       
4 &$\mathbf{H}^{T}_{1,1}\mathbf{F}_{1}\mathbf{H}_{1,2}$
&$8N^{2}_{T}N_{i}+4N_{T}N^{2}_{i}+2N_{T}N_{i}$&
696\\
\hline 5 &$\mathbf{H}^{T}_{2,1}\mathbf{F}_{2}\mathbf{H}_{2,2}$
&$8N^{2}_{T}N_{i}+4N_{T}N^{2}_{i}+2N_{T}N_{i}$ & 696\\ \hline 6 &
$\mathbf{\widehat{C}}_{1}$
 &$2K(32N^{2}_{T}N_{i}+8N_{T}N_{i}+2N^{2}_{T}-4N_{i}+3N_{T})$ & 14856 \\ \hline 7 &$(\mathbf{\Xi}^{H}_{i}\mathbf{\Xi}_{i})^{-1}$ &$K(\frac{14}{3}N^{3}_{T}-2N^{2}_{T}+N_{T})$ &
 2826
 \\
\hline 8 & $\mathrm{det}\mathbf{B}^{2}_{1}$
 &$4K(N^{3}_{T}+N^{2}_{T}+2N_{T})$ & 3168
\\ \hline Total &  & & 33530 \\
\hline
\end{tabular}
\end{table*}

\begin{table*}[tbp]
\centering  
\caption{Computational complexity of the nonregenerative MIMO
relay system \cite{b19}.}
\begin{tabular}{| c | c | c | c |} \hline 
Step   & Operations  & FLOPS & Case: $(2,2,2)\times 6$ \\ \hline  
1
&$\mathbf{U}^{a}_{i}\mathbf{\Sigma}^{a}_{i}\mathbf{\Lambda}^{aH}_{i}$
&$8K(4N^{2}_{T}N_{i}+8N_{T}N^{2}_{i}+9N^{3}_{i})$ & 13248
\\ \hline        
2
&$\mathbf{U}^{a}_{j}\mathbf{\Sigma}^{a}_{j}\mathbf{\Lambda}^{aH}_{i}$
&$ 8K(4N^{2}_{T}N_{i}+8N_{T}N^{2}_{i}+9N^{3}_{i})$ & 13248
\\ \hline       
3 &$\mathbf{H}^{H}_{i}\mathbf{H}_{i}$ &$4KN_{i}N_{T}(N_{i}+1)$&
432\\
\hline 4 &$\mathbf{H}^{H}_{j}\mathbf{H}_{j}$
&$4KN_{i}N_{T}(N_{i}+1) $ & 432\\ \hline
5 &$\mathbf{H}^{H}_{i}[\sigma^{2}_{1}\sigma^{2}_{2}(\mathbf{H}_{j}\mathbf{F})^{H}\mathbf{H}_{j}\mathbf{F}+\mathbf{I}]^{-1}\mathbf{H}_{i}$ &$2K(N^{3}_{i}+8N_{i}N^{2}_{T}+4N^{2}_{i}N_{T}+2N_{i}N_{T}-N^{2}_{i}+N_{i})$ & 4212 \\
\hline 6 &$\mathbf{V}_{A}\mathbf{\Lambda}_{A}\mathbf{V}^{H}_{A}$
 &$8K(4N^{2}_{T}N_{i}+8N_{T}N^{2}_{i}+9N^{3}_{i}+\frac{1}{2}N_{i}) $ & 13272 \\ \hline 7 &$\mathrm{diag}(\widetilde{\mathbf{G}})$ &$K[4N_{i}N_{T}(N_{i}+1)+2N^{3}_{i}-2N^{2}_{i}+N_{i}]$ & 462

\\ \hline Total &  & & 45306 \\
\hline
\end{tabular}
\end{table*}

\begin{table*}[tbp]
\centering  
\caption{Computational complexity of the conventional RBD
\cite{b18}.}
\begin{tabular}{| c | c | c | c |} \hline 
Step   & Operations  & FLOPS & Case: $(2,2,2)\times 6$ \\ \hline  
1
&$\mathbf{U}^{a}_{i}\mathbf{\Sigma}^{a}_{i}\mathbf{\Lambda}^{aH}_{i}$
&$32K(N_{T}\overline{N}^{2}_{i}+2\overline{N}^{3}_{i})$ & 21504
\\ \hline        
2
&$\left((\mathbf{\Sigma}^{a}_{i})^{T}\mathbf{\Sigma}^{a}_{i}+\rho^{2}\mathbf{I}\right)^{-1/2}$
&$ K(18N_{T}N^{2}_{i}-2N^{2}_{i})$ & 336
\\ \hline       
3 &$\mathbf{V}^{a}_{i}\mathbf{D}^{a}_{i}$ &$8KN^{3}_{T}$&
5184\\
\hline 4 &$\mathbf{H}_{i}\mathbf{P}^{a}_{i}$ &$K(8N_{T}N^{2}_{i}-2N^{2}_{i}) $ & 552\\
\hline
5 &$\mathbf{U}^{b}_{i}\mathbf{\Sigma}^{b}_{i}\mathbf{V}^{bH}_{i}$ &$64K(\frac{9}{8}N^{3}_{i}+N_{T}N^{2}_{i}+\frac{1}{2}N^{2}_{T}N_{i})$ & 13248 \\
\hline Total &  & & 40824  \\
\hline
\end{tabular}
\end{table*}

\begin{table*}[tbp]
\centering  
\caption{Computational complexity of the CD-BD algorithm
\cite{c2}.}
\begin{tabular}{| c | c | c | c |} \hline 
Step   & Operations  & FLOPS & Case: $(2,2,2)\times 6$ \\ \hline  
1
&$\mathbf{U}^{H}_{i,1}\mathbf{\Sigma}_{i,1}\mathbf{\Lambda}_{i,1}$
&$8K(4N^{2}_{T}N_{i}+8N_{T}N^{2}_{i}+9N^{3}_{i})$ & 13248
\\ \hline        
2
&$\mathbf{\Lambda}^{H}_{i,2}\mathbf{\Sigma}_{i,2}\mathbf{U}_{i,2}$
&$8K(4N^{2}_{T}N_{i}+8N_{T}N^{2}_{i}+9N^{3}_{i})$ & 13248
\\ \hline       
3
&$\mathbf{H}_{i,2}\mathbf{W}\mathbf{H}_{i,1}$&$K[8N_{i}N^{2}_{T}-2N_{i}N_{T}+4N_{i}N_{T}\times(N_{i}+1)]$&
2088\\
\hline 4 &$\mathbf{L}^{H}_{i}\mathbf{L}_{i}$ &$2K(N_{i}+2N_{T}N_{i}\times(N_{i}+1)+4N^{3}_{i}/3) $ & 508\\
\hline 5 &$\mathbf{H}^{\dag}_{mse}$
&$4N^{3}_{R}/3+12N^{2}_{R}N_{T}-2N^{2}_{R}-2N_{T}N_{R}$ & 2736 \\
\hline 6 &$\mathbf{H}_{i,i}\mathbf{V}^{a}_{i}\mathbf{V}^{b}_{i}$
&$8K[4N_{T}N^{2}_{i}-4N^{3}_{i}/3+N^{2}_{i}(N_{i}+1)]$ & 2336
\\\hline
7 &$(\mathbf{Q}_{i}\mathbf{Q}^{H}_{i}+\sigma^{2}_{i}\Psi_{i})^{-1}$ &$K[4N_{R}N_{i}\times(N_{i}+1)+3N_{i}+2N^{3}_{i}-2N^{2}_{i}]$ & 474 \\
\hline Total &  & & 34638  \\
\hline
\end{tabular}
\end{table*}

\section{Simulation Results}
In this section, we study the performance of the proposed QL-QR
algorithm for two-way MIMO relay networks. All the simulations are
performed on the assumption that all the channels are the Rayleigh
fading channel and they are independently generated following
$\sim{CN(0,1)}$. The noise variances $\sigma^{2}_{i}$ are equally
given as $\sigma^{2}$. The total relay power constraint can be
written as
\begin{eqnarray}\label{Constraint}
P_{R_{1}}+P_{R_{2}}=aP_{R}+(1-a)P_{R}=P_{R},
\end{eqnarray}
where $a\in[0,1]$ is an auxiliary value as well as a power
allocation coefficient between two relay nodes. All the simulation
results are averaged over 1000 channel trials.

\begin{figure}
\begin{center}
\includegraphics [width=92mm,height=75mm]{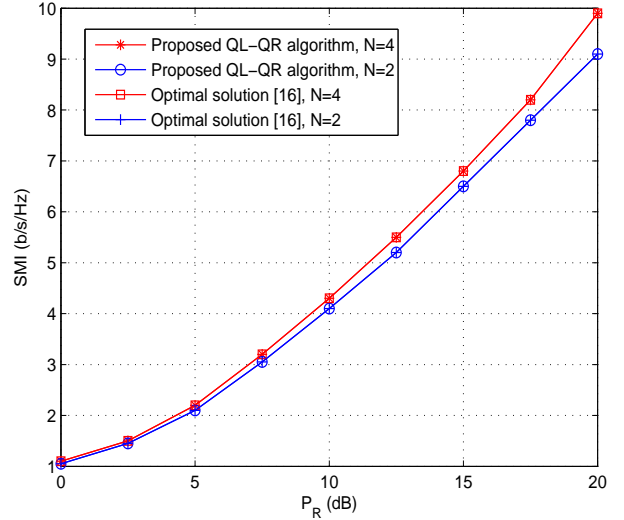}\\
\caption{\label{7} The achieved SMI for $N=4, 2$.}
\end{center}
\end{figure}

In Fig. \ref{7}, we compare the sum mutual information (SMI) of
various MU-MIMO schemes where full CSI is known at each node. We
set $P_{1}=P_{2}=10~\mathrm{dB}$, $M=1$, and an equal power budget
for the two relays ($a=0.5$ is assumed). The negative SMI is
adopted in \cite{b15} which can be defined as
\begin{eqnarray}\label{SMI}
\mathbf{MI}_{sum}=\mathrm{log}_{2}\left|\mathbf{MSE}_{1}\right|+\mathrm{log}_{2}\left|\mathbf{MSE}_{2}\right|.
\end{eqnarray}
In our proposed method, the SMI  shown in the simulation results
is calculated as $-2~\mathrm{log}_{2}|\mathbf{B}^{2}_{i}|$ by
using \eqref{Choleskey}, \eqref{det1}, and \eqref{det2}. It can be
observed that the proposed QL-QR algorithm has the same SMI
performance as an optimal solution in \cite{b15}.

\begin{figure}
\begin{center}
\includegraphics [width=92mm,height=75mm]{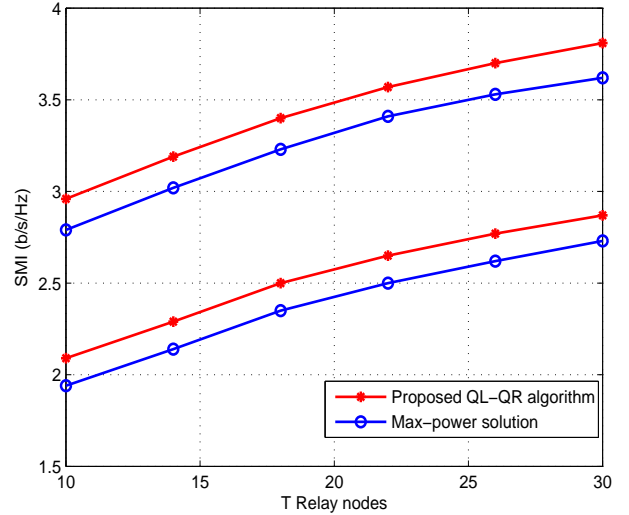}\\
\caption{\label{8} The SMI versus the number of the relays $T$.}
\end{center}
\end{figure}

Figure \ref{8} shows the performance of our proposed SMI
performance versus the number of the relays, $T$ which is even. We
consider a practical scenario with different relay power
constraints. We set $P_{R}=30~\mathrm{dB}$ and $a=0.5$. It is
clear that, for different values of $P_{1}$ and $P_{2}$, a
solution of our proposed QL-QR algorithm shows better performance
than a max-power solution.

\begin{figure}
\begin{center}
\includegraphics [width=92mm,height=75mm]{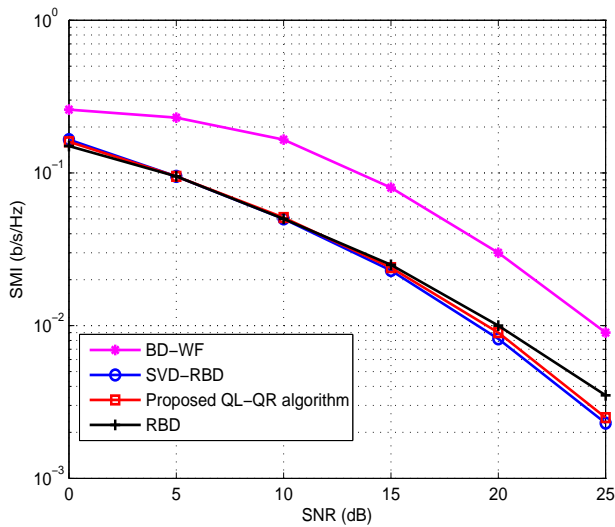}\\
\caption{\label{9} BER performance on the Rayleigh fading channel
.}
\end{center}
\end{figure}

Figure \ref{9} exhibits the BER performance of the BD water
filling, the RBD, the SVD-RBD, and our proposed QL-QR method,
where the QPSK modulation is made use of. As pointed out in
\cite{e1}, the BER performance for a MIMO precoding system is
actually determined by the energy of the transmitted signal. To
simplify our discussion, we assume $a=0$. In the RBD,
$\mathrm{det}\left(\overline{\mathbf{H}}\overline{\mathbf{H}}^{H}\right)=\prod^{m}_{i=1}\lambda^{2}_{i}$,
where $\overline{\mathbf{H}}\in{\mathbb{C}^{N\times M}}$, for
$M<N$, is an equivalent channel matrix with its eigenvalues
$\lambda_{i}$. In our proposed QL-QR method, for source node
$\mathbf{S}_{1}$, we have
$\mathrm{det}\left(\mathbf{\widehat{H}}_{1}\mathbf{\widehat{H}}^{H}_{1}\right)=\prod^{m}_{i=1}\varsigma^{2}_{1}$.
Under the stipulaton that $\mathrm{det}\mathbf{F}_{D,i}=1$, we are
able to easily obtain $\lambda_{i}=\varsigma_{1}$. Therefore, our
proposed QL-QR method has the same BER performance as the SVD-RBD
method.

\section{Conclusions}

This paper studies joint optimization problem of an AF based on
the MIMO TWRC, where two source nodes exchange their messages with
two relay nodes. A relay filter has been designed, which is able
to efficiently join the source and the relay nodes. Our main
contribution is that the optimal beamforming vectors can
efficiently be computed using determinant maximization techniques
through an iterative QL-QR algorithm based on a MSE balancing
method. Our proposed QL-QR algorithm can significantly reduce the
computational complexity and has the equivalent BER performance to
the SVD-BD algorithm.

\section*{Acknowledgment}
The authors would like to thank the anonymous reviewer for their
great constructive comments that improved this paper.

\end{document}